\documentclass[aps,preprint,showpacs,floatfix]{revtex4}
\usepackage{graphicx,bm,amsmath,dcolumn}

 \usepackage{graphicx}
 \RequirePackage{color}

 \definecolor{MyDarkGreen}{rgb}{0.02,0.60,0.06}

\graphicspath{{figures}}
\begin{document}
\title{A generalised formulation of the Laplacian approach to resistor networks}

\author{N.Sh. Izmailian}
\email{izmail@yerphi.am; ab5223@coventry.ac.uk}
\affiliation{Applied Mathematics Research Centre, Coventry University, Coventry CV1 5FB, UK}
\affiliation{Yerevan Physics Institute, Alikhanian Brothers 2, 375036 Yerevan, Armenia}

\author{R. Kenna}
\email{r.kenna@coventry.ac.uk}
\affiliation{Applied Mathematics Research Centre, Coventry University, Coventry CV1 5FB, UK}

\date{\today}

\begin{abstract}
An analytic approach is presented to developing exact expressions for the two-point resistance between arbitrary nodes on certain non-regular resistor networks.
This generalises previous approaches, which only deliver results for networks of more regular geometry.
The new approach exploits the second minor of the Laplacian matrix associated with the given network to obtain the  resistance in terms its eigenvalues and eigenvectors.
The method is illustrated by application to the resistor network on the  globe lattice, for which the resistance between two arbitrary nodes is obtained in the form of single summation.
\end{abstract}


\maketitle
\section{Introduction}
\label{Introduction}
The calculation of the resistance between two arbitrary nodes in resistor networks is the classic problem in electric circuit theory and was first studied by Kirchhoff in 1847 \cite{kirch}.
The problem has attracted the interest of numerous physicists over many years because it is intrinsically connected to a wide range of other physics problems, including   random walks \cite{random1,random2,random3,random4}, first-passage processes \cite{passage}, lattice Green's functions \cite{green1,green2,history} and classical transport in disordered media \cite{media1,media2,media3}.
The resistance $R_{i,j}$ between two  nodes $i$ and $j$ can be considered as a metric called ``resistance distance'' \cite{Klein}.
If there are many (few) paths between the two nodes, the resistance $R_{i,j}$  is small (large).
The total resistance distance of a graph, also called the ``Kirchhoff index'' \cite{Klein} (i.e., the sum of resistance distances between all pairs of nodes) is related to the network criticality \cite{network}, which characterises its robustness.

In the past, resistance-computation studies have focused mainly on infinite lattices \cite{history,green1,green2,asad1,asad2,asad3}.
Recently, attention has shifted  to the study of resistance on finite networks, as these are the configurations of relevance to real life.
For this reason, there has been a surge of research activity in recent years.
In 2004 Wu \cite{wu} derived a compact expression for the resistance between  two arbitrary nodes for finite, regular lattice networks in terms of
the associated Laplacian.
That approach, however, requires a complete knowledge of the eigenvalues and eigenvectors of the Laplacian.
This is straightforward to obtain for regular lattices in any dimensions, since the Laplacian for $d$-dimensional regular lattices can be represented as the sum of $d$ one-dimensional Laplacians, with known eigenvalues and eigenvectors.
However, the approach cannot readily deal with non-regular lattices.
For this reason, Izmailian, Kenna and Wu extended the approach to enable derivation of a closed-form expression for the resistance between two arbitrary nodes for finite networks in terms of the eigenvalues and eigenvectors of the {\emph{first minor}} of the Laplacian \cite{ikw}.
The new approach  has been applied to the cobweb and fan resistor networks \cite{ikw,ik}.
An alternative recent approach to the calculation of two-point resistances on distance-regular networks was based on the stratification of the network and the associated Stieltjes function \cite{jafar1}.
Still another approach  of computing resistances by using a method of direct summation has been developed in \cite{tan0,tan2globe}.

In this paper the recent approaches \cite{wu,ikw} are further generalized to compute two-point resistances based on the eigenvalues and eigenvectors of the {\emph{second}} minor of the Laplacian associated with the network.
The generalized approach is illustrated by application to the resistor network with configuration of a globe. In particular, the resistance between two arbitrary nodes on such a network is determined in the form of single summation, allowing determination to arbitrary precision.

\section{Resistor networks}
\label{Resistors}

Let us consider a resistor network consisting of $T$ nodes and let $r_{i,j}=r_{j,i}$ be the resistance  of the resistor connecting nodes $i$ and $j$. The resistance between arbitrary nodes $\alpha$ and $\beta$  can be written as \cite{wu}
\begin{equation}
R_{\alpha,\beta}=\sum_{i=2}^{T}\frac{\left|\psi_{i\alpha}-\psi_{i\beta}\right|^2}{\lambda_i},
\label{Rab}
\end{equation}
where $\lambda_i$ are nonzero eigenvalues with orthonormal eigenvectors  $\Psi_{i}=(\psi_{i1},\psi_{i2},...,\psi_{iT})$ of the Laplacian ${\bf L}_T$ of that network.

Determining the eigenvalues and eigenvectors of the Laplacian is usually a very difficult problem.
One way of approaching it is to reduce the original problem to that of 1D Laplacians with appropriate boundary conditions.
For regular rectangular lattices, this can be achieved in any dimension in a straightforward manner.
The eigenvalues and eigenvectors of the Laplacian for the 1D lattice with various boundary conditions are then easy to calculate and they given in Appendix 1.

Let us now consider, for example, the Laplacian for regular, two-dimensional,  rectangular lattices.
Denote by ${\bf L}_N^{\rm{free}}$, ${\bf L}_N^{\rm{per}}$, ${\bf L}_N^{\rm{DN}}$ and ${\bf L}_N^{\rm{DD}}$  the Laplacian of a 1D lattice with free, periodic, Dirichlet-Neumann and Dirichlet-Dirichlet boundary conditions, respectively and let $I_K$ be the $K \times K$ identity matrix.
Then, the 2D Laplacian of the resistor network ${\bf L}_{M \times N}$  consisting of a $M \times N$ rectangular lattice with free, cylindrical and toroidal boundary conditions and with resistors $r$ and $s$ in the two directions, can be expressed through the Laplacians of the 1D lattices  as \cite{wu}
\begin{eqnarray}
{\bf L}_{M \times N}^{\rm{free}}&=&r^{-1}{\bf L}_N^{\rm{free}}\otimes {\bf I}_M+s^{-1}{\bf I}_N \otimes {\bf L}_M^{\rm{free}},
\label{LapFree}\\
{\bf L}_{M \times N}^{\rm{cylinder}}&=&r^{-1}{\bf L}_N^{\rm{free}}\otimes {\bf I}_M+s^{-1}{\bf I}_N \otimes {\bf L}_M^{\rm{per}},
\label{LapCyl}\\
{\bf L}_{M \times N}^{\rm{torus}}&=&r^{-1}{\bf L}_N^{\rm{per}}\otimes {\bf I}_M+s^{-1}{\bf I}_N \otimes {\bf L}_M^{\rm{per}}.
\label{LapTor}
\end{eqnarray}
Thus, the 2D Laplacian ${\bf L}_{M \times N}$ can be diagonalize in the two subspaces separately, yielding eigenvalues and eigenvectors
\begin{eqnarray}
\lambda_{m,n}&=&r^{-1}\lambda_m+s^{-1}\lambda_n
\label{eigenvalue}\\
\psi_{(m,n);(x,y)}&=&\psi_{mx}^{(M)}\psi_{ny}^{(N)}
\end{eqnarray}
where $\lambda_m, \lambda_n$ and $\psi_{mx}^{(M)},\psi_{ny}^{(N)}$ are eigenvalues and eigenvectors of the appropriate 1D Laplacian.

But for non-regular lattices, such as the cobweb and fan networks consisting of $M \times N + 1$ sites, it is impossible to express the 2D Laplacian ${\bf L}_{M \times N+1}$ of the network through the Laplacians of such 1D lattices.
This means that it is difficult to apply Wu's method \cite{wu}.
Instead, on can apply the method of Izmailian, Kenna and Wu (``IKW method'') \cite{ikw} to compute resistance by using eigenvalues and eigenvectors of the {\emph{first minor}} of the 2D Laplacian.
Indeed, the first minor ${\bf \Delta}_{M \times N}$ of the 2D Laplacian ${\bf L}_{M \times N+1}$ can be reduced to the Laplacian of a 1D lattice and can be written as \cite{ikw,ik}
\begin{eqnarray}
{\bf \Delta}_{M \times N}^{\rm{cobweb}}&=&r^{-1}{\bf L}_N^{\rm{per}}\otimes {\bf I}_M+s^{-1}{\bf I}_N \otimes {\bf L}_M^{\rm{DN}},\label{D1cobweb}\\
{\bf \Delta}_{M \times N}^{\rm{fan}}&=&r^{-1}{\bf L}_N^{\rm{free}}\otimes {\bf I}_M+s^{-1}{\bf I}_N \otimes {\bf L}_M^{\rm{DN}}.
\label{D1fan}
\end{eqnarray}
Then the resistance between nodes $\alpha$ and $\beta$ can be written as \cite{ikw}
\begin{equation}
R_{\alpha,\beta}=\sum_{i=1}^{M\times N} \frac{\left|\psi_{i\alpha}-\psi_{i\beta}\right|^2}{\lambda_i},
\label{Rab}
\end{equation}
where $\lambda_i$ are eigenvalues with orthonormal eigenvectors  $\Psi_{i}=(\psi_{i1},\psi_{i2},...,\psi_{iT})$ of the minor ${\bf \Delta}_{M\times N}$. Note, that all eigenvalues of the minor ${\bf \Delta}_{M\times N}$ have nonzero value.

Therefore, to compute resistances  on regular $M \times N$ rectangular lattices with free, cylindrical and toroidal boundary conditions, one can use the Wu method \cite{wu}.
To compute resistances  on non-regular rectangular lattices, such as the cobweb and fan networks one can use IKW method \cite{ikw}.
The main difference between these two approaches is that in the Wu method one expresses the resistance through the eigenvalues and eigenvectors of the full Laplacian of the network, while in the IKW method the resistance is expressed through the eigenvalues and eigenvectors of the first minor of the Laplacian of the network.

There are, however, other non-regular rectangular lattices, such as the globe network comprising $M \times N + 2$ sites, for which it is impossible to express the Laplacian ${\bf L}_{M \times N+2}$ or the first minor of the Laplacian through Laplacians 1D lattices.
It is therefore difficult to apply either the Wu \cite{wu} or IKW \cite{ikw} methods to calculate resistances between nodes for such a network,
However, in this circumstance, the second  minor ${\bf {\cal L}}_{M \times N}$ of the Laplacian ${\bf L}_{M \times N+2}$ can be written as
\begin{eqnarray}
{\bf {\cal L}}_{M \times N}^{\rm{globe}}&=&r^{-1}{\bf L}_N^{\rm{per}}\otimes {\bf I}_M+s^{-1}{\bf I}_N \otimes {\bf L}_M^{\rm{DD}}.
\label{D2globe}
\end{eqnarray}
In what follows we shall show how to calculate resistances $R_{\alpha,\beta}$ on the globe  network using the modified Laplacian approach, i.e., by expressing the resistance through eigenvalues and eigenvectors of the second minor of the Laplacian of the network.
Extension to networks of similar geometries is possible in a straightforward manner.

\section{Modified Laplacian approach}
\label{Resistors2}

Let us consider a network, in which the total number of nodes is $M  N +2$.
Let us denote the nodes by the index $i$, wherein $i$ takes values $0, 1, 2, ..., M  N +1$. Denote the electric potential at the $i$th node by $V_i$ and the current flowing into the network at the $i$th node by $I_i$.
We write Kirchhoff's law as
\begin{equation}
\sum_{j=0}^{M N+1}c_{i,j}(V_i-V_j)=I_i, \qquad i=0,1,2,...,M N, M N+1,
\label{Kirchhoff}
\end{equation}
with the constraint
\begin{equation}
\sum_{i=0}^{MN+1}I_i=0.
\label{constraint}
\end{equation}
Here $c_{i,j}=c_{j,i}$ is the conductance, which can be expressed through the resistance of the resistor connecting nodes $i$ and $j$ $r_{i,j}=r_{j,i}$ as
\begin{equation}
c_{i,j}=r_{i,j}^{-1}.
\label{condactance}
\end{equation}
Eq.~(\ref{Kirchhoff}) can be written in matrix form as
\begin{equation}
{\bf L} {\vec V} = {\vec I},
\label{matrix0}
\end{equation}
where ${\bf L}$ is the Laplacian of the system
$$
{\bf L}=\left( \begin{array}{ccccccc}
c_0 & -c_{0,1} & -c_{0,2} &\ldots &-c_{0,M N}&-c_{0,M N+1} \\
-c_{1,0} & c_1 & -c_{1,2}&\ldots  &-c_{1,M N}&-c_{1,M N+1}\\
\vdots & \vdots & \vdots &\ddots &\vdots & \vdots & \vdots \\
-c_{M N,0}&-c_{M N,1}&-c_{M N,2}&\ldots&c_{M N}&-c_{M N, M N+1}\\
-c_{M N+1,0}&-c_{M N+1,1}&-c_{M N+1,2}&\ldots&-c_{M N+1,M N}&c_{M N+1}
\end{array} \right),
$$
and $c_i$ is given by
\begin{equation}
c_i=\sum_{j=0 \above0pt  j\neq i}^{M N+1}c_{i,j} \qquad i=0,1,2,...,M N+1.
\label{ci}
\end{equation}
Here, ${\vec V}$ and ${\vec I}$ are vectors
$$
{\vec V}=\left( \begin{array}{ccccccc}
V_0\\
V_1\\
V_2\\
\vdots \\
V_{MN}\\
V_{MN+1}
\end{array} \right),
\qquad
{\vec I}=\left( \begin{array}{ccccccc}
I_0\\
I_1\\
I_2\\
\vdots \\
I_{MN}\\
I_{MN+1}
\end{array} \right).
$$
Under the constraint (\ref{constraint}) we actually have only $M N+1$ independent equations in Eq. (\ref{Kirchhoff}).
Without loss of generality, therefore, we choose to delete the equation numbered $i=M N +1$ and choose the potential at node $M N+1$ to be zero: $V_{MN+1}=0$.
Then the $M N +2$ equations in (\ref{Kirchhoff}) are reduced to the set of independent $M N +1$ equations
\begin{equation}
\sum_{j=0}^{M N+1}c_{ij}(V_i-V_j)=I_i, \qquad i=0,1,2,...,MN.
\label{Kirchhoff1}
\end{equation}
To this point, we have followed the ref.~\cite{ikw}.
Next we partition the set of equations (\ref{Kirchhoff1}) into two parts.
The first is a single equation and the second is a set of $M N$ equations, {\emph{viz.}}
\begin{eqnarray}
\sum_{j=1}^{M N+1}c_{0j}(V_0-V_j)&=&I_0, \label{single}\\
\sum_{j=0}^{M N+1}c_{ij}(V_i-V_j)&=&I_i, \qquad i=1,2,...,MN.
\label{Kirchhoff2}
\end{eqnarray}
The set of equations (\ref{Kirchhoff2}) can be written in the matrix form as
\begin{equation}
{\bf {\cal L}} {\vec {\cal V}} = {\vec {\cal I}},
\label{matrix}
\end{equation}
where ${\bf {\cal L}}$ is the second minor of the Laplacian ${\bf L}$ and is given by
$$
{\bf {\cal L}}=\left( \begin{array}{ccccccc}
c_1 & -c_{1,2} & -c_{1,3} &\ldots &-c_{1,M N} \\
-c_{2,1} & c_2 & -c_{2,3}&\ldots  &-c_{2,M N}\\
\vdots & \vdots & \vdots &\ddots &\vdots & \vdots & \vdots \\
-c_{M N,1}&-c_{M N,2}&-c_{M N,3}&\ldots&c_{M N}
\end{array} \right),
$$
and ${\vec {\cal V}}$ and ${\vec {\cal I}}$ are vectors, which are now given by
$$
{\vec {\cal V}}=\left( \begin{array}{ccccccc}
V_1\\
V_2\\
V_3\\
\vdots \\
V_{MN}
\end{array} \right),
\qquad
{\vec {\cal I}}=\left( \begin{array}{ccccccc}
{\cal I}_1\\
{\cal I}_2\\
{\cal I}_3\\
\vdots \\
{\cal I}_{MN}
\end{array} \right).
$$
Here ${\cal I}_i=I_i+c_{i,0}V_0$, for $i=1,2,3,...,MN$.

Eq.~(\ref{matrix}) can now be straightforwardly solved for ${\vec V}$ since ${\bf {\cal L}}^{-1}$ is not singular. Multiplying from the left by ${\bf {\cal L}}^{-1}$, we obtain the solution ${\vec V}={\bf {\cal L}}^{-1} {\vec {\cal I}}$. Explicitly, this reads
\begin{eqnarray}
V_i&=&\sum_{j=1}^{M N} {\cal L}^{-1}_{i,j} {\cal I}_j =V_0\sum_{j=1}^{MN}{\cal L}^{-1}_{i,j}c_{j,0}+\sum_{j=1}^{M N} {\cal L}^{-1}_{i,j} I_j, \qquad i=1,2,...,M N,
\label{Vi1}
\end{eqnarray}
where ${\cal L}^{-1}_{i,j}$  is the $(i,j)$th elements of the inverse matrix ${\bf {\cal L}^{-1}}$.

Since we choose  the potential at node $M N+1$ to be zero, $V_{MN+1}=0$, Eq.~(\ref{single}) can be transformed as
\begin{eqnarray}
c_0 V_0&=&I_0+\sum_{i=1}^{MN}c_{0,i}V_i \label{single1}\\
&=&I_0+\sum_{i=1}^{MN}\sum_{j=1}^{MN}{\cal L}_{ij}^{-1}c_{0,i}I_j+V_0\sum_{i=1}^{MN}\sum_{j=1}^{MN}{\cal L}_{ij}^{-1}c_{j,0}c_{0,i}.
\label{single2}
\end{eqnarray}
From Eq. (\ref{single2}) we can find $V_0$,
\begin{equation}
V_0=\frac{I_0+\sum_{i=1}^{MN}\sum_{j=1}^{MN}{\cal L}_{ij}^{-1}c_{0,i}I_j}{c_0-\sum_{i=1}^{MN}\sum_{j=1}^{MN}{\cal L}_{ij}^{-1}c_{j,0}c_{0,i}}.
\label{V00}
\end{equation}
Plugging this expression for $V_0$ back to  Eq. (\ref{Vi1}) we obtain for $V_k$ the expression
\begin{eqnarray}
V_k&=&\frac{I_0+\sum_{i=1}^{MN}\sum_{j=1}^{MN}{\cal L}_{ij}^{-1}c_{0,i}I_j}{c_0-\sum_{i=1}^{MN}\sum_{j=1}^{MN}{\cal L}_{ij}^{-1}c_{j,0}c_{0,i}}\sum_{j=1}^{MN}{\cal L}^{-1}_{k,j}c_{j,0}+\sum_{j=1}^{M N} {\cal L}^{-1}_{k,j} I_j, \qquad k=1,2,...,M N.
\label{Vi2}
\end{eqnarray}
Thus Eqs. (\ref{V00}) and (\ref{Vi2}) give us expressions for $V_i$ ($i=0,1,2,...,MN$) in terms of the elements of the inverse matrix ${\bf {\cal L}^{-1}}$.

To compute the resistance $R_{\alpha \beta}$ between arbitrary  two nodes $\alpha$ and $\beta$, we connect $\alpha$ and $\beta$ to an external battery and measure the current $I$ going through the battery with no other nodes are connected to external sources. Let the potentials at the two nodes be, respectively, $V_{\alpha}$ and $V_{\beta}$. Then, by Ohm's law, the desired resistance is
\begin{equation}
R_{\alpha \beta} = \frac{V_{\alpha}-V_{\beta}}{I}.
\label{resistance}
\end{equation}
The computation of $R_{\alpha \beta}$ is now reduced to solving Eqs. (\ref{single}) and (\ref{Kirchhoff2}) for $V_{\alpha}$ and $V_{\beta}$ with the current given by
\begin{equation}
I_j = I (\delta_{j\alpha}-\delta_{j\beta}).
\label{current}
\end{equation}
Combining Eqs. (\ref{resistance}) and (\ref{current}) with Eq. (\ref{Vi2}) we obtain the resistance between any two nodes $\alpha$ and $\beta$ other than the zeroth node ($\alpha = 0$ or $\beta = 0$) as
\begin{equation}
R_{\alpha,\beta}={\cal L}^{-1}_{\alpha,\alpha} + {\cal L}^{-1}_{\beta,\beta}  - {\cal L}^{-1}_{\alpha,\beta}- {\cal L}^{-1}_{\beta \alpha}+\frac{\sum_{i=1}^{MN}\left({\cal L}_{i,\alpha}^{-1}-{\cal L}_{i,\beta}^{-1}\right)c_{0,i}\sum_{j=1}^{MN}\left({\cal L}_{\alpha,j}^{-1}-{\cal L}_{\beta,j}^{-1}\right)c_{j,0}}{c_0-\sum_{i=1}^{MN}\sum_{j=1}^{MN}{\cal L}_{ij}^{-1}c_{j,0}c_{0,i}}.
\label{resistorab}
\end{equation}
Here we have used the fact that, in the case of any two nodes $\alpha$ and $\beta$ other than $0$, the current $I_0=0$, as  follow from Eq. (\ref{current}).
Up to now all our considerations have been quite general.
We next illustrate the method by application to the globe resistor network.

\section{Application of new Approach to the Globe  Resistor Network}
\label{Resistors1}

\begin{figure}[tbp]
\includegraphics[width=0.3\textwidth]{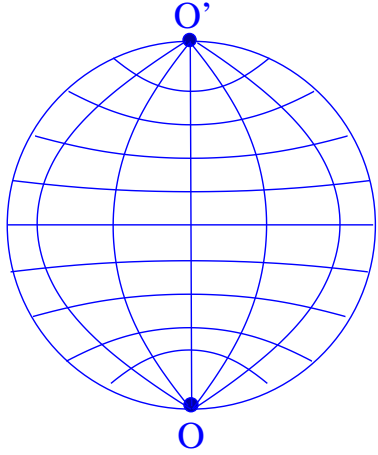}
\caption{Illustration of a spherical $M \times N$ + 2 globe network.
Here there are $M=9$ latitudinal rows and $N= 7$ longitudinal ones.
Periodic boundary conditions are imposed in the latitudinal direction.
Each site on the bottom row is  connected to the site labelled $O$ and each site on the top row is connected to $O^\prime$. } \label{fig1}
\end{figure}

Here we consider an example of a network for which the Wu and IKW approaches fail.
One such network is that with geometry of a  globe, an example of which is illustrated in Figure~1.
The  network ${\cal L}_{\rm globe}$ is an $M \times N$ rectangular lattice with periodic boundary condition in one direction and with nodes on each of the two boundaries in the other direction connected to two single nodes.
Topologically ${\cal L}_{\rm globe}$ is in the form of a globe consisting of N longitudinal lines and M lines of latitude, with two poles $O$ and $O'$.
The total number of node in globe  network is $M  N +2$.
Bonds in longitudinal and latitudinal directions have resistances $s$ and $r$, respectively. The elements $c_{0,i}=c_{i,0}$ of the globe  network  have the following values
\begin{eqnarray}
c_{0,i}=c_{i,0}&=&s^{-1} \qquad \mbox{for} \quad i=1,2,3,...,N , \nonumber\\
c_{0,i}=c_{i,0}&=&0 \qquad \mbox{for} \quad i=N+1,N+2,N+3,...,MN , \nonumber
\end{eqnarray}
and $c_0$ is given by
\begin{equation}
c_0=\sum_{j=1}^{MN+1}c_{0,j}=N s^{-1} .
\end{equation}
Then Eq.~(\ref{resistorab}) can be transformed as
\begin{equation}
R_{\alpha,\beta}={\cal L}^{-1}_{\alpha,\alpha} + {\cal L}^{-1}_{\beta,\beta}  - {\cal L}^{-1}_{\alpha,\beta}- {\cal L}^{-1}_{\beta \alpha}+\frac{\sum_{i=1}^{N}\sum_{j=1}^{N}\left({\cal L}_{i,\alpha}^{-1}-{\cal L}_{i,\beta}^{-1}\right)\left({\cal L}_{\alpha,j}^{-1}-{\cal L}_{\beta,j}^{-1}\right)}{N s-\sum_{i=1}^{N}\sum_{j=1}^{N}{\cal L}_{ij}^{-1}}.
\label{resistorab1}
\end{equation}
Here, ${\cal L}^{-1}_{i,j}$  is the $(i,j)$th element of the inverse matrix ${\bf {\cal L}^{-1}}$ which  is given by
\begin{equation}
{\cal L}^{-1}_{i,j}=\sum_{k=1}^{MN}\frac{\psi_{k,i}\psi_{k,j}^{*}}{\Lambda_k},
\label{Delta1}
\end{equation}
in which $\Lambda_k$ and $\psi_{k,i}$ are the eigenvalues and eigenvectors of the second minor ${\bf {\cal L}}$ of the Laplacian.

The second minor of the Laplacian for the globe network is given by
\begin{eqnarray}
{\bf {\cal L}}_{M \times N}^{globe}&=&r^{-1}{\bf L}_N^{\rm{per}}\otimes {\bf I}_M+s^{-1}{\bf I}_N \otimes {\bf L}_M^{\rm{DD}},
\label{MtimesNGlobe}
\end{eqnarray}
where ${\bf L}_N^{\rm{per}}$ and ${\bf L}_N^{\rm{DD}}$ are the Laplacians of 1D lattices with periodic and Dirichlet-Dirichlet  boundary conditions, respectively, and ${\bf I}_M$ and ${\bf I}_N$ are identity matrices.
The eigenvalues and eigenvectors of ${\bf L}_N^{\rm per}$ and ${\bf L}_M^{\rm{DD}}$ are known (see Appendix A). This leads to the following eigenvalues and eigenvectors for  the second minor of the Laplacian for the globe network:
\begin{eqnarray}
\Lambda_k \equiv
\Lambda_{m,n}&=&2r^{-1}(1-\cos{2\theta_n})+2s^{-1}(1-\cos{2\varphi_m}),
\label{Lambdamn}\\
\psi_{k,i} \equiv \psi_{(m,n);(x,y)}&=&\sqrt{\frac{2}{N(M+1)}}\exp\left(2 i x \theta_n\right)\sin(2 y \varphi_m),\label{varphiki}
\end{eqnarray}
where $\theta_n$ and $\varphi_m$ are given by Eqs. (\ref{thetan}) and (\ref{varphin}) respectively.

We next calculate the two double sums in Eq. (\ref{resistorab1}), namely $S_1=\sum_{i=1}^{N}\sum_{j=1}^{N}{\cal L}_{ij}^{-1}$ and $S_2=\sum_{i=1}^{N}\sum_{j=1}^{N}\left({\cal L}_{i,\alpha}^{-1}-{\cal L}_{i,\beta}^{-1}\right)\left({\cal L}_{\alpha,j}^{-1}-{\cal L}_{\beta,j}^{-1}\right)$.
Our objective is not to actually determine the two double summations exactly.
Rather it is to reduce each of them a single summation, because a single summation allows one to determine the resistance to arbitrary precision numerically.

Let us start first with double sum $S_1$.
Since the coordinates of the nodes from $i=1,2,...,N$ are given by $(x,y)$, where the $x$ coordinate takes values from 1 to N ($x=1,2,3,...,N$) and the $y$ coordinate takes value 1 ($y=1$), the first sum can be written as
\begin{equation}
S_1=\sum_{i=1}^{N}\sum_{j=1}^{N}{\cal L}_{ij}^{-1}=\sum_{i=1}^{N}\sum_{j=1}^{N}\sum_{k=1}^{MN}\frac{\psi_{k,i}\psi_{k,j}^*}{\Lambda_k}=
\sum_{x_1=1}^{N}\sum_{x_2=1}^{N}\sum_{m=0}^{M-1}\sum_{n=0}^{N-1}\frac{\psi_{(m,n);(x_1,1)}
\psi_{(m,n);(x_2,1)}^*}{\Lambda_k}.
\end{equation}

In the case of the globe network we should use $\Lambda_k$ and $\psi_{k,i}$ given by Eqs. (\ref{Lambdamn}) and (\ref{varphiki}) and for $S_1=S_1^{globe}$ we obtain
\begin{equation}
S_1^{\rm{globe}}=\frac{1}{2N(M+1)}\sum_{x_1=1}^{N}\sum_{x_2=1}^{N}\sum_{n=0}^{N-1}\sum_{m=0}^{M-1}\frac{\exp{(2 i (x_1-x_2)\theta_n)} \sin^2(2\varphi_m)}{r^{-1}\sin^2\theta_n+s^{-1}\sin^2\varphi_m}
\label{S1G}.
\end{equation}
Using the  identities
\begin{eqnarray}
\sum_{x=1}^N \exp{(2 i x \theta_n)}&=&\sum_{x=1}^N \exp{(2 \pi i x n/N)}=N\delta_{n,0}\label{identity1},\\
\sum_{m=0}^{M-1}\cos^2\varphi_m&=&\sum_{m=0}^{M-1}\cos^2\left(\frac{\pi(m+1)}{2(M+1)}\right)=\frac{M}{2}\label{identity2},
\end{eqnarray}
This reduces to the simple expression
\begin{equation}
S_1^{\rm{globe}}=\frac{M N s}{M+1}\label{S1GH}.
\end{equation}

The second double sum $S_2$  can be written as a product of two single sum, namely
$$
S_2=\sum_{i=1}^{N}\left({\cal L}_{i,\alpha}^{-1}-{\cal L}_{i,\beta}^{-1}\right) \times \sum_{j=1}^{N}\left({\cal L}_{\alpha,j}^{-1}-{\cal L}_{\beta,j}^{-1}\right)
$$
Let us choose the coordinates of nodes $\alpha$ and $\beta$ as $\alpha=(x_1,y_1)$ and $\beta=(x_2,y_2)$.
Then the first single sum in the case of the globe network can be written as
\begin{eqnarray}
\lefteqn{\sum_{i=1}^{N}\left({\cal L}_{i,\alpha}^{-1}-{\cal L}_{i,\beta}^{-1}\right)}
\\
&=&\sum_{x=1}^{N}\sum_{m=0}^{M-1}\sum_{n=0}^{N-1}\frac{\psi_{(m,n);(x,1)}
\left(\psi_{(m,n);(x_1,y_1)}^*-\psi_{(m,n);(x_2,y_2)}^*\right)}{\Lambda_{m,n}}\nonumber\\
&=&\frac{1}{2N(M+1)}\sum_{x=1}^{N}\sum_{m=0}^{M-1}\sum_{n=0}^{N-1}
\frac{e^{2 i x \theta_n}\sin(2\varphi_m)\left(e^{-2 i x_1 \theta_n}\sin(2y_1\varphi_m)-e^{-2 i x_2 \theta_n}\sin(2y_2\varphi_m)\right)}{r^{-1}\sin^2\theta_n+s^{-1}\sin^2\varphi_m}\nonumber\\
&=&\frac{1}{2(M+1)}\sum_{m=0}^{M-1}\sum_{n=0}^{N-1}
\frac{\delta_{n,0}\sin(2\varphi_m)\left(e^{-2 i x_1 \theta_n}\sin(2y_1\varphi_m)-e^{-2 i x_2 \theta_n}\sin(2y_2\varphi_m)\right)}{r^{-1}\sin^2\theta_n+s^{-1}\sin^2\varphi_m}\nonumber\\
&=&\frac{s}{M+1}\sum_{m=0}^{M-1}\cot\varphi_m\left(\sin(2y_1\varphi_m)-\sin(2y_2\varphi_m)\right)\nonumber\\
&=&\frac{s}{M+1}(y_2-y_1).\label{sum1G3}
\end{eqnarray}
Here we have use the  identity
\begin{equation}
\sum_{m=0}^{M-1}\cot\varphi_m \sin(2y\varphi_m)=\sum_{m=0}^{M-1}\cot\frac{\pi (m+1)}{2(M+1)}\sin\frac{\pi (m+1) y}{M+1}=M+1-y ,\label{identity4}
\end{equation}
which holds for integer values of $y$.
The second single sum can be obtain in the similar manner:
\begin{eqnarray}
\sum_{j=1}^{N}\left({\cal L}_{\alpha,j}^{-1}-{\cal L}_{\beta,j}^{-1}\right)&=&
\frac{s}{M+1}(y_2-y_1)\label{sum1G4}
\end{eqnarray}
Thus for the second double sum $S_2$ in the case of globe network we obtain
\begin{equation}
S_2=S_2^{globe}=\frac{s^2(y_2-y_1)^2}{(M+1)^2}.
\end{equation}

Substituting Eq. (\ref{Delta1}) into Eq. (\ref{resistorab1}) and plugging the double sums $S_1$ and $S_2$ back to Eq. (\ref{resistorab1}) we finally obtain for the resistance between two nodes $\alpha=(x_1,y_1)$ and $\beta=(x_2,y_2)$ of the globe network the  expression
\begin{equation}
R_{\alpha,\beta}=\frac{s(y_2-y_1)^2}{N(M+1)}+\sum_{m=0}^{M-1}\sum_{n=0}^{N-1}
\frac{\left|\psi_{(m,n);(x_1,y_1)}-\psi_{(m,n);(x_2,y_2)}\right|^2}{\Lambda_{m,n}}.
\label{resistorabfinal}
\end{equation}

From Eq. (\ref{resistorabfinal}) and using expressions  (\ref{Lambdamn}) and (\ref{varphiki}) for the eigenvalues and eigenvectors of the second minor of the Laplacian of the globe network, we obtain for the resistance $R^{\rm{globe}}_{\alpha,\beta}$ between
two nodes at $\alpha = \{x_1, y_1\}$ and $\beta = \{x_2, y_2\}$,
\begin{eqnarray}
R ^{\rm{globe}}_{\alpha,\beta}&=&\frac{s(y_2-y_1)^2}{N(M+1)}+\sum_{m=0}^{M-1}\sum_{n=0}^{N-1}
\frac{\left|\phi_{(m,n);(x_1,y_1)}-\phi_{(m,n);(x_2,y_2)}\right|^2}
{\Lambda_{m,n}}\nonumber\\
&=&\frac{s(y_2-y_1)^2}{N(M+1)}+\frac{r}{N(M+1)}\sum_{m=0}^{M-1}\sum_{n=0}^{N-1}
\frac{S_1^2+S_2^2-2S_1S_2\cos[2(x_1-x_2)\theta_n]}
{(1-\cos{2\theta_n})+ h (1-\cos{2\varphi_m})},
\label{R12}
\end{eqnarray}
where
\begin{equation}
h=r/s, \qquad S_1=\sin\left(2y_1\varphi_{m}\right), \qquad S_2=\sin\left(2y_2\varphi_{m}\right). \nonumber
\end{equation}

It is convenient to introduce the quantity $\varLambda(\varphi_m)$ by writing
\begin{equation}
1 + h (1-\cos \varphi_m ) = {\rm ch} 2 \varLambda (\varphi_m) \nonumber
\end{equation}
or,
\begin{equation}
{\rm sh} \varLambda(\varphi_m) = \sqrt h \sin \varphi_m\, .
\end{equation}
We can then carry out the summation over $n$ in Eq.~(\ref{R12}) by using the summation identity
\begin{equation}
\frac 1 N \sum_{n=0}^{N-1} \frac {\cos (2 \ell \theta_n)}
 {{\rm ch} 2 \varLambda - \cos 2\theta_n   } = \, \frac{{\rm ch}[(N -2 \ell)\varLambda)]}
{{\rm sh} (2\varLambda ){\rm sh}(N \varLambda)},
\label{sumidentity}
\end{equation}
with  $\ell = 0, \, |x_1-x_2|$, to obtain
\begin{eqnarray}
R ^{\rm{globe}}_{\alpha,\beta}&=& \frac{r} {M+1}
\sum_{m=0}^{M-1}\frac{S_1^2+S_2^2-2S_1 S_2
{\rm ch}\big[2|x_1-x_2|\,\varLambda_m\big]}
{{\rm sh}(2\,\varLambda_m)}\,\coth ( N\,\varLambda_m ) \nonumber\\
&+&\frac{r} {M+1}  \sum_{m=0}^{M-1}
\frac{ 2S_1S_2{\rm sh}\big[2|x_1-x_2|\,\varLambda_m \big]}
{{\rm sh} (2\,\varLambda_m)}+\frac{s(y_2-y_1)^2}{N(M+1)}, \label{R12y}
\end{eqnarray}
where $\varLambda_m = \varLambda (\varphi_m)$.
This is our desired expression -- the resistance as a single summation.
Note, that in Ref.~\cite{wu} the exact expression for the two-point resistance on regular lattices was obtained in the form of a double summation only.
One requires the summation identities given by Eq. (\ref{sumidentity}) to reduce the  expression  to the form of single summation.

In the special case of $x_1 = x_2=x$, i.e., two nodes in the same $y$ column at $y_1$ and $y_2$, Eq.~(\ref{R12y}) reduces to
\begin{equation}
R ^{\rm{globe}}(x_1=x_2=x) = \frac{s(y_2-y_1)^2}{N(M+1)}+\frac{r} {M+1}\sum_{m=0}^{M-1}\frac {\coth (N \varLambda_m)}
  {{\rm sh} (2\varLambda_m)} \big[\sin (2y_1 \varphi_m) -\sin (2y_2 \varphi_m)\big]^2.
\label{x1x2}
\end{equation}
similarly, in the special case of $ y_1 = y_2=y$, wherein two nodes are in the same $x$ row at $x_1$ and $x_2$, Re.~(\ref{R12y}) reduces to
\begin{equation}
R ^{\rm{globe}}(y_1=y_2=y) = \frac{4r} {M+1}\sum_{m=0}^{M-1}
\frac { {\rm sh} \big[|x_1-x_2| \varLambda_m\big]
        {\rm sh} \big[ \big(N-|x_1-x_2| \big) \varLambda_m \big]  }
      {{\rm sh} (2 \varLambda_m ) {\rm sh} (N \varLambda_m) }  \sin^2 (2 y \varphi_m). \label{y1y2}
\end{equation}

Note that our results (\ref{R12y})-(\ref{y1y2}) coincides with corresponding results of \cite{tan2globe}, obtained by alternative approach of computing resistances by using a method of direct summation.

\section{Summary}
\label{Summary}

We have revisited the problem of the evaluation of two-point resistances
in a resistor network using the Laplacian approach considered in \cite{wu,ikw}.
We reformulated the problem in terms of the eigenvalues and eigenfunctions of the second minor of the Laplacian $\cal L$.
We showed that this strategy can deliver solutions in circumstances outside the reach of previous Laplacian based approaches.
As an example, the new formulation is applied to the globe resistor network: a cylindrical lattice with sites on end boundaries connected to two external common nodes $O$ and $O'$.  Our analysis leads to an exact expression (\ref{R12y}) for the resistance between arbitrary two nodes on the globe network, from which numerical results may be determined to arbitrary precision.

\section{Acknowledgments}
\label{Acknowledgments}
The work was supported by a Marie Curie IIF (Project no. 300206-RAVEN)
and IRSES (Projects no. 295302-SPIDER and 612707-DIONICOS) within 7th European Community Framework
Programme and by the grant of the Science Committee of the Ministry of Science and
Education of the Republic of Armenia under contract 13-1C080. In addition, we would like to thank John Essam for discussions.


\appendix

\section{1D Laplacians: eigenvalues and eigenvectors}
\label{1DLaplac}

On a domain of rectangular shape the $d$-dimensional Laplacian ${\bf L^{(d)}}$ is a sum of $d$ one-dimensional Laplacians ${\bf L^{(1)}}$ if the boundary conditions in  one direction do not depend on the coordinates in other directions. If we call $\alpha_1,\alpha_2,...,\alpha_d$ the boundary conditions in the $1,2,...,d$ directions, the d-dimension Laplacian ${\bf L^{(d)}}_{\alpha_1,\alpha_2,...,\alpha_d}$ can be written as
\begin{equation}
{\bf L^{(d)}}_{\alpha_1,\alpha_2,...,\alpha_d}= L^{(1)}_{\alpha_1} \otimes {\bf I}\otimes...\otimes {\bf I}+{\bf I} \otimes L^{(1)}_{\alpha_2} \otimes ...\otimes {\bf I}+...+{\bf I} \otimes .... \otimes {\bf I} \otimes L^{(1)}_{\alpha_d}.
\nonumber
\end{equation}
If ${\mathbf f_{\alpha_i}(x)}$ is an eigenfunction of ${\bf
L_{\alpha_i}^{(1)}}$, with eigenvalue ${\bf \lambda_{\alpha_i}}$, then the
product $\prod_{i=1}^d {\bf f_{\alpha_i}(x)}$ is an eigenvalue of
${\bf L_{\alpha_1,\alpha_2,...,\alpha_d}^{(d)}}$ with eigenvalue $\sum_{i=1}^d{\bf
\lambda_{\alpha_i}}$.
Thus the spectra of one-dimensional Laplacians is all one needs to
diagonalize higher dimensional versions.

Let us consider a 1D lattice with $N$ sites, with coordinate
${\bf x}$ between $1$ and $N$.
The lattice version of the Laplacian is given by
\begin{eqnarray}
{\bf L} f(x)= 2\; f(x) - f(x + 1) - f (x - 1),\nonumber
\end{eqnarray}
where $f(x)$ is the vector
$$
{\bf f(x)}=\left( \begin{array}{ccccccc}
f(1)\\
f(2)\\
f(3)\\
\vdots \\
f(N-1)\\
f(N)
\end{array} \right).
$$
Now we are going to solve the following equation:
\begin{equation}
{\bf L} \; f(x) = \lambda \; f(x), \hspace{1cm} x = 1,2,..., N .\nonumber
\end{equation}
In the lattice version,
\begin{equation}
2\; f(x) - f(x + 1) - f (x - 1) = \lambda \; f(x), \hspace{1cm} x = 1,2,..., N \nonumber
\end{equation}
where $f(x)$ is an eigenfunction and $\lambda$ is an eigenvalue. The solution is given by
\begin{equation}
f(x) = a \; e^{i \theta x} + b \; e^{- i \theta x}, \hspace{1cm}
\lambda = 2 - 2 \cos{\theta},  \nonumber
\end{equation}
in which the coefficients $a$, $b$ and $\theta$ are fixed by specific boundary conditions and by normalization conditions for $f(x)$: $\sum_{x=1}^{N}f(x)^*f(x)=1$

Let us now consider the following boundary conditions of the 1D lattice

\subsection{Periodic: $f(1)=f(N+1)$}

The Laplacian  for periodic boundary conditions is given by
$$
{\bf L}_N^{\rm{per}}=\left( \begin{array}{ccccccc}
2 & -1 & 0 &\ldots &0&0&-1 \\
-1 & 2 & -1&\ldots &0 &0&0\\
\vdots & \vdots & \vdots &\ddots &\vdots & \vdots & \vdots \\
0&0&0&\ldots&-1&2&-1\\
-1&0&0&\ldots&0&-1&2
\end{array} \right).
$$
The eigenvectors $f_n(x)$ and eigenvalues $\lambda_n$ are given by
\begin{eqnarray}
f_n(x) &=& \sqrt{\frac{1}{N}}\; {\rm exp}(2\,i  \theta_n \, x ), \qquad x=1,2,...,N \nonumber\\
\lambda_n &=& 2 - 2\cos (2 \, \theta_n) , \qquad n=0,1,...,N-1,
\nonumber
\end{eqnarray}
where $\theta_n$ is given by
\begin{equation}
\theta_n=\frac{\pi n}{N}.
\label{thetan}
\end{equation}

\subsection{Free (Neumann-Neumann): $f(0)=f(1)$ and $f(N)=f(N+1)$}

The Laplacian  for free (Neumann-Neumann) boundary conditions is given by
$$
{\bf L}_N^{\rm{free}}=\left( \begin{array}{ccccccc}
1 & -1 & 0 &\ldots &0&0&-1 \\
-1 & 2 & -1&\ldots &0 &0&0\\
\vdots & \vdots & \vdots &\ddots &\vdots & \vdots & \vdots \\
0&0&0&\ldots&-1&2&-1\\
-1&0&0&\ldots&0&-1&1
\end{array} \right).
$$
The eigenvectors $f_n(x)$ and eigenvalues $\lambda_n$ are given by
\begin{eqnarray}
f_n(x)  &=& \frac{1}{\sqrt{N}} \qquad \mbox{for} \quad n=0 \nonumber\\
    &=& \sqrt{\frac{2}{N}}\;\cos((x-1/2)\,\theta_n), \qquad \mbox{for} \quad n=1,2,...,N-1 \nonumber\\
\lambda_n &=& 2 - 2\cos (\theta_n) , \qquad n=0,1,...,N-1,
\nonumber
\end{eqnarray}
where $\theta_n$ is given by Eq. (\ref{thetan}).

\subsection{Dirichlet-Dirichlet: $f(0)=f(N+1)=0$}

The Laplacian  for Dirichlet-Dirichlet boundary conditions is given by
$$
{\bf L}_N^{\rm{(DD)}}=\left( \begin{array}{ccccccc}
2 & -1 & 0 &\ldots &0&0&0 \\
-1 & 2 & -1&\ldots &0 &0&0\\
\vdots & \vdots & \vdots &\ddots &\vdots & \vdots & \vdots \\
0&0&0&\ldots&-1&2&-1\\
0&0&0&\ldots&0&-1&2
\end{array} \right),
$$
The eigenvectors $f_n(x)$ and eigenvalues $\lambda_n$ are given by
and
\begin{eqnarray}
f_n(x) &=& \sqrt{\frac{2}{N+1}}\sin (2 \, x \varphi_n )\nonumber\\
\Lambda_n &=& 2 - 2\cos (2 \varphi_n) , \qquad n=0,...,N-1,
\label{lambda}
\end{eqnarray}
where $\varphi_n$ is given by
\begin{equation}
\varphi_n=\frac{\pi (n+1)}{2(N+1)}
\label{varphin}
\end{equation}

\subsection{Dirichlet-Neumann: $f(0)=0$ and $f(N)=f(N+1)$}

If one chooses, for instance, a left Dirichlet and a right
Neumann boundary, then the Laplacian is given by
$$
{\bf L}_N^{\rm{(DN)}}=\left( \begin{array}{ccccccc}
2 & -1 & 0 &\ldots &0&0&0 \\
-1 & 2 & -1&\ldots &0 &0&0\\
\vdots & \vdots & \vdots &\ddots &\vdots & \vdots & \vdots \\
0&0&0&\ldots&-1&2&-1\\
0&0&0&\ldots&0&-1&1
\end{array} \right).
$$
The eigenvectors $f_n(x)$ and eigenvalues $\lambda_n$ are given by
and
\begin{eqnarray}
f_n(x) &=& \frac{2}{\sqrt{2N+1}}\sin (2 \, x \phi_n )\, \nonumber\\
\lambda_n &=& 2 - 2\cos (2 \phi_n) , \qquad n=0,...,N-1,
\label{lambda}
\end{eqnarray}
where $\phi_n$ is given by
\begin{equation}
\phi_n=\frac{\pi \left(n+\frac{1}{2}\right)}{2N+1}.
\label{varphin}
\end{equation}


\begin{thebibliography}{99}


\bibitem{kirch} G. Kirchhoff, Ann. Phys. Chem. {\bf 72}, 497 (1847).

\bibitem{random1} P. G. Doyle and J. L. Snell, {\it Random Walks and Electric
Networks}, (The Carus Mathematical Monograph, series 22, The
Mathematical Association of America, USA, 1984) pp. 83-149.

\bibitem{random2} M. Jeng, Am. J. Phys. {\bf 68}, 37 (2000).

\bibitem{random3} N. Chair, Annals of Physics {\bf 327}, 3116 (2012).

\bibitem{random4} N. Chair, Annals of Physics {\bf 341}, 56 (2014).

\bibitem{passage} S. Redner, {\it A Guide to First-Passage Processes} (Cambridge University
Press, Cambridge, 2001)

\bibitem{green1} S. Katsura, T. Morita, S. Inawashiro, T. Horiguchi and Y. Abe, J. Math. Phys. {\bf 12}, 892 (1971).

\bibitem{history} J. Cserti, Am. J. Phys {\bf 68}, 896  (2000).

\bibitem{green2} J. Cserti, G. Szechenyi, G. David, J. Phys. A: Math. Theor. {\bf 44}, 215201  (2011).

\bibitem{media1} S. Kirkpatrick, Rev. Mod. Phys. {\bf 45}, 574 (1973).

\bibitem{media2} B. Derrida, J. Vannimenus, J. Phys. A {\bf 15}, L557 (1982).

\bibitem{media3} A. B. Harris, T. C. Lubensky, Phys. Rev. B {\bf 35}, 6964 (1987).

\bibitem{Klein} D. J. Klein, M. Randic, J. Math. Chem. {\bf 12}, 81 (1993).

\bibitem{network} A. Tizghadam, A. Leon-Garcia, IEEE Network {\bf 24}, 10 (2010).

\bibitem{asad1} J. H. Asad, A. Sakaji, R. S. Hijjawi, J. M. Khalifeh, Eur. J. Phys. B {\bf 52}, 365 (2006).

\bibitem{asad2} J. H. Asad, A. A. Diab, R. S. Hijjawi, J. M. Khalifeh, Eur. Phys. J. Plus PLUS  {\bf 128}, 2 (2013).

\bibitem{asad3} J. H. Asad, J. Stat. Phys. {\bf 50}, 1177 (2013).


\bibitem{wu} F. Y. Wu, J. Phys. A: Math. Gen. {\bf 37} 6653 (2004).

\bibitem{ikw} N. S. Izmailian, R. Kenna and F. Y. Wu, J. Phys. A: Math. Gen. {\bf 47} 035003 (2014).

\bibitem{ik} N. S. Izmailian and R. Kenna, {\it The two-point resistance of fan networks}, preprint arXiv:1401.4463

\bibitem{jafar1} M. A. Jafarizadeh, R. Sufiani and S. Jafarizadeh, J. Phys. A: Math. Theor. {\bf 40}, 4949 (2007).

\bibitem{tan0} Z. Z. Tan, {\it Resistor network models} (in Chinese), Xindian University of Science and Technology Press, Xian, China (2011).

\bibitem{tan2globe} Z. Z. Tan, J. W. Essam and F. Y. Wu, {\it The two-point resistance of a cobweb with a superconducting boundary}, preprint arXiv:1404.2350

\end{thebibliography}
\end{document}